# Weak antilocalization in $Cd_3As_2$ thin films

Bo Zhao[1], Peihong Cheng[2], Haiyang Pan[1], Shuai Zhang[1], Baigeng Wang[1], Guanghou Wang[1], Faxian Xiu[2]*, Fengqi Song[1]*

[1]National Laboratory of Solid State Microstructures, Collaborative Innovation Center of Advanced Microstructures, and College of Physics, Nanjing University, Nanjing, 210093, P. R. China

[2]State Key Laboratory of Surface Physics and Department of Physics, Collaborative Innovation Center of Advanced Microstructures, Fudan University, Shanghai 200433, P. R. China

*Correspondence authors: F. X. and F. S. (E-mails: faxian@fudan.edu.cn and songfengqi@nju.edu.cn)

**ABSTRACT**

Recently, it has been theoretically predicted that $Cd_3As_2$ is a three dimensional Dirac material, a new topological phase discovered after topological insulators, which exhibits a linear energy dispersion in the bulk with massless Dirac fermions. Here, we report on the low-temperature magnetoresistance measurements on a ~50nm-thick $Cd_3As_2$ film. The weak antilocalization under perpendicular magnetic field is discussed based on the two-dimensional Hikami-Larkin-Nagaoka (HLN) theory. The electron-electron interaction is addressed as the source of the dephasing based on the temperature-dependent scaling behavior. The weak antilocalization can be also observed while the magnetic field is parallel to the electric field due to the strong interaction between the different conductance channels in this quasi-two-dimensional film.

**Introduction**

Dirac materials, with linear band dispersion in low-energy excitation such as graphene and topological insulators, have been receiving increasing attention owing to the possibility of being a new candidate for next-generation electronic and spintronic devices[1,2]. Recently, theory predicts that $Cd_3As_2$[3]and $Na_3Bi$[4] are the three dimensional (3D) Dirac materials, soon after which, they have been experimentally demonstrated by angle-resolved photoemission spectroscopy (ARPES)[5-7], scanning tunneling microscope (STM)[8] and electrical transport measurements[9]. Remarkably, various novel topological phases, such as Weyl semimetals, topological insulators and topological superconductors can be obtained from 3D Dirac materials by breaking the time reversal symmetry or inversion symmetry[10,11]. More importantly, electrical transport measurements of $Cd_3As_2$ bulk crystals exhibit many novel phenomenona, such as high mobility, giant and linear magnetoresistance (MR), non-trivial quantum oscillations and Landau level splitting under magnetic fields[8,9,12,13]. Another important phenomenon is the negative MR which confirms the existence of chirality in Weyl fermions that has been observed in $Cd_3As_2$ nanowires[14] and nanoplates[15]. And in the systems of $Bi_{0.97}Sb_{0.03}$[16], TaAs[17,18], $ZrTe_5$[19] and $Na_3Bi$[20], the chiral anomaly induced negative MR has also been observed.

In low-dimensional structures, studies of electrical transport have revealed several quantum interference (QI) phenomena at low temperature, including weak localization (WL) or weak antilocalization (WAL) and universal conductance fluctuations (UCF). The physical properties of some materials can be determined from these QI phenomena including the temperature dependence of the resistance, Hall effect, and MR. In 2D films, these contributions are logarithmic in temperature and have nontrivial dependence on the magnetic field. In perpendicular and parallel fields, MR always shows different behavior since orbital QI effects are sensitive to field orientation [21-23]. These techniques are well established but continue to be useful in the study of electrical transport in a variety of systems [22-24]. WAL phenomenon is always observed in Dirac materials, such as topological insulators and graphene without inter-valley scattering, as an important consequence of spin-momentum locking and the full suppression of backscattering, resulting in a relative $\pi$ Berry phase acquired by electrons executing time-reversed paths[25,26]. Recently, the WAL effect has also been observed in 3D Dirac semimetal $Cd_3As_2$[27,28], 3D Weyl semimetal TaAs[17,18] and $Bi_{0.97}Sb_{0.03}$[16]. In addition, the WAL effect by theoretical description using the Feynman diagram shows its origin from the inter-valley scattering as described in two-dimensional (2D) Dirac materials such as graphene and topological insulators[29].

However, the WAL effect is always absent in 2D films when applied parallel magnetic field due to the suppressed interference in a closed electron path. As an effective tool to investigate the scattering ratio in parallel field, the WAL effect is indispensable. In this work, we study the magnetoresistance of ~50nm-thick $Cd_3As_2$ films. We have found a sharp cusp around $B = 0$ which is stable when the magnetic field is perpendicular to the film (B⊥E) or parallel to electric field (B∥E). We take into account 2D HLN formula in the presence of spin-orbit coupling to discuss the WAL effect in B⊥E while we explore the origin of the WAL effect in B∥E, where a negative MR is also observed.

**Results and Discussion**

**Figure 1** shows the temperature dependence of the resistance $R_{xx}$ of sample CA2. With the temperature decreasing, $R_{xx}$ shows an increasing behavior over the temperature range which is different from metallic bulk materials. This insulating behavior is also reported in low-dimensional $Cd_3As_2$ materials [27,28,30] and bulk materials under high pressure [31,32]. The data can be fitted to the Arrhenius formula of $R_{xx} \sim exp(E_g/2k_BT)$ at temperatures from 100K to 200K, where $E_g$ is the bandgap, $k_B$ is the Boltzmann constant and $T$ is the measurement temperature. We obtain the band gap $E_g = 21.9 meV$, which is reasonable with the value for the $Cd_3As_2$ thin film of this thickness [3,28]. When the temperature reaches <100K, the resistance deviates from the Arrhenius formula and shows weak temperature dependence below 20K. The sharp increase below 4K indicates a contribution from electron-electron scattering in the presence of disorder in the 2D films at low temperatures [33].

**Figure 2** (a) and (b) show the MR under applied magnetic fields $B \perp E$ and $B // E$ respectively at various temperatures. The MR defined as:

$$MR(\%) = \frac{R_{xx}(B) - R_{xx}(0)}{R_{xx}(0)} \times 100\%$$

For the perpendicular field of 9T, the MR is around 100% with a weak temperature dependent below 150K. This is different from the giant MR in bulk samples [34]. Around $B = 0T$, sharp cusps are observed clearly and gradually weaken with increasing temperature. This cusp is ascribed to the quantum interference phenomenon, weak antilocalization, which is also observed in B∥E shown in **Figure 2 (b)**. However, the MR shows a different behavior in B∥E contrast with that in B⊥E. It decreases with increasing applied magnetic field, showing a negative MR.

Before discussing the magneto-transport properties of the $Cd_3As_2$ films, we first investigate their dimensionality. The film can be treated as a 2D system for the thickness $t$ smaller than the appropriate physical length scales. As shown in **Table I**, the electronic mean-free path $\ell_e$ for three samples are obtained with the 2D formula $\ell_e = \hbar\sqrt{2\pi n_e}\mu/e$, where $\hbar$ is the Planck constant, $n_e$ is the 2D carrier density obtained from the Hall effect measurement, μ is the carrier mobility extracted from the electronic conductivity formula $\sigma = n_e e \mu$. It is clear that the mean-free path $\ell_e$ is longer than the thickness $t$=50nm for sample CA1 and CA3 and less than the thickness for sample CA2. So the classical diffusive transport is quasi-2D and the treatment before is reasonable. For QI effects, the relevant length scale is the dephasing length $L_\varphi$. In nonmagnetic weak disorder systems, the dephasing length is always dominated by inelastic scattering, such as electron-phonon and electron-electron scattering. The electron-phonon scattering with a strong temperature-dependent is always suppressed with decreasing temperature. Thus the $L_\varphi$ increases with the temperature decreasing. In the measurement temperature from 100K to 2K, $L_\varphi > t$ (based on the analysis below) indicates a 2D behavior in the QI range. So we restrict our analysis to this quasi-2D limit.

In **Figure 3** and **4**, the magnetoconductivity are treated with $\Delta\sigma(B) = \sigma(B) - \sigma(0)$ at various temperatures, where $\sigma = (L/W)(1/R_{xx}(B))$, $L$ and $W$ are the length and width of the sample respectively, $R_{xx}(B)$ is the resistance under applied magnetic field $B \perp E$ or $B // E$.

In 2D systems with an applied perpendicular magnetic field, Hikami, Larkin and Nagaoka (HLN) first described in the presence of spin-orbit coupling for the temperature dependent WAL correction to the conductance[35]. Besides, the background with parabolic conductance contribution is also considered[36]. The correction formula can be written as:

$$\Delta\sigma_\perp(B) \cong \alpha \frac{-e^2}{2\pi^2\hbar}\left[\Psi\left(\frac{1}{2} + \frac{B_\varphi}{B}\right) - \ln\left(\frac{B_\varphi}{B}\right)\right] + c \cdot B^2 \qquad (1)$$

where $\Psi(\chi)$ is the digamma function, the characteristic field $B_\varphi = \hbar/(4eL_\varphi^2)$, and $L_\varphi$ is effective dephasing length. The parameter α takes values of 1/2 and -1 respectively for weak antilocalization and weak localization. The fitting curves for samples CA1 are shown in **Figure 3(a)** with red solid curves. The parameter α, shown in **Figure 3(c)**, with a value of 0.2 at 2K that is less than it proposed by theory. The temperature dependent of dephasing length $L_\varphi$ is shown in **Figure 3(b)**. It is clear that the $L_\varphi$ decreases from 513nm to 110nm with the temperature increasing from 2K to 100K. The extracted $L_\varphi$ values are larger than the thickness of $Cd_3As_2$ films, justifying the 2D WAL characteristics in our samples.

The parameter α reflects the number of independent conduction channels in the film[25,37]. For the 2D electron gas (2DEG) with parabolic dispersion, the magnetoconductivity can transform from WL to WAL (α from -1 to 0.5) as a function of the strength of scattering off spin-orbit impurities. However, for the massless Dirac fermions in 3D topological insulators, the WAL (α=0.5) always exist for every value of the spin-orbit disorder [38]. Besides, due to the coexistence of topologically trivial 2DEG and topological surface state, the α<0.5 is observed in 3D topological insulator[39]. For the negative magnetoconductivity in 3D Dirac semimetal, the WAL in low field and short-range scattering in high field are also discussed [29,40]. So the several independent conduction channels may coexist with each other in this $Cd_3As_2$ films with different strength of spin-orbit scattering. It induce α<0.5.

In the low temperature regime, the dephasing mechanism for 2D system can induce a power-law rule of $L_\varphi^{-2} \sim T^2$ due to the electron-phonon scattering and $L_\varphi^{-2} \sim T$ because of the quasielastic Nyquist electron-electron scattering process[41]. Here we proposed the electron-electron interaction and the saturated dephasing mechanism in the $Cd_3As_2$ film and it can be fit with the formula $L_\varphi^{-2} \sim L_0^{-2} + AT$ [41,42], where $L_0$ represents the zero-temperature dephasing length. It is sensitive to the impurities and surface scattering. The last term $AT$ is the contribution from electron-electron interaction. As shown in **Figure 3(b)**, it results in a perfect fitting with the parameter $L_0 = 626.2nm$. Above all, the electron-electron interaction and the surface scattering or impurities may exist in this film[41].

**Figure 4(a)** shows the WAL effect of the sample CA1 under $B /\!/ E$. It clearly displays evident cusp around $B = 0T$ with the 2D character confirmed by $L_\varphi \gg t$. In thin films, Al'tshuler and Aronov (AA) described firstly the quantum corrections to the conductivity when applied magnetic field in the plane of the film [43] and a similar suppression of QI phenomenon as in the perpendicular field case. The formula for this correction under $B /\!/ E$ can be written as:[23,43]

$$\Delta\sigma_{\|}(B) \cong \alpha \frac{-e^2}{2\pi^2\hbar} \ln\left(1 + \beta \frac{et^2}{4\hbar B_{\varphi/\!/}} B^2\right) \quad (2)$$

Where $\Delta\sigma_{/\!/}(B) = \sigma_{/\!/}(B) - \sigma_{/\!/}(0)$. $t = 50nm$ is the film thickness. The parameter $\beta$ is related to the ratio of mean free path and film thickness. The meaning of the parameter α is described before. However, the value of α=0.16 in $B /\!/ E$ at 2K is similar to α=0.2 in $E \perp B$, and with a similar temperature dependence in **Figure 3(c)**. So we assume that the dephasing length in $B /\!/ E$ is equal to the one under $B \perp E$ ($B_{\varphi\|} = B_{\varphi\perp}$) [22]. Thus, the $\beta$ vs temperature can be obtained. The $\beta$ of three samples are shown in **Figure 4(b)**. For CA1, the value of $\beta$ is close to 1 at 2K and decreases to 0.56 when the temperature increases to 30K. In this sample, the mean free path $\ell_e = 64nm$ is larger than the film thickness. So this result shows a good agreement with the value of $0 < \beta < 1$ for $L_\varphi \gg \ell_e > t$ in bilayer system, such as double quantum wells and topological insulators films with two separate surface states [22,44]. For CA2 and CA3, it shows a smaller value $\beta$ than that in CA1, but it also shows a larger value than the theory proposed with $\beta < 1/3$ for $\ell_e < t$ in single layer systems, as shown in **Figure 4 (b)**.

Although the origin of this obvious WAL effect in this 2D film in B∥E is still a puzzle, a physical origin is proposed to explain this: There are at least two 2D conductance channels contribution in $Cd_3As_2$ films. This field introduces an additional phase for the electron which moves in one channel, then tunnels into another channel, moves there, and finally tunnels back and returns to the initial point.

For B∥E, the MC is upturned with increasing field away from WAL effect range. The experimental observation of this crossover is always attributed to the translation from WAL to WL. The weak localization effect due to the QI is always suppressed by thermal average at higher temperatures. However, for the negative MR in B∥E in Figure 2(b), it is rather robust and survives at higher temperature. However, the WL also observed in graphene at room temperature[45]. Besides, this upturn is enhanced in applied magnetic field B∥E.

**Conclusion**

In summary, we have measured the resistance and magnetoresistance of $Cd_3As_2$ films with thickness of ~50nm under applied magnetic field B⊥E or B∥E. An insulating temperature dependent resistance with a band gap $E_g = 21.89 meV$ is observed in $Cd_3As_2$ films. Positive MR sharp cusp around $B = 0T$ under both applied field were measured, which can be satisfactorily described by existing 2D WAL theory. The electron-electron scattering is suggested as a source of the dephasing mechanism in this $Cd_3As_2$ film. Under B∥E, the WAL effect is also clearly observed and proposed as the strong coupling between the different conductance channels in this quasi-two-dimensional film. The negative MR is observed in B//E in $Cd_3As_2$ films. But the exact origin of this negative MR needs further work.

**Methods**

The ~50nm-thick $Cd_3As_2$ thin films are grown in a molecular beam epitaxy system. The devices are fabricated with standard Hall bar geometry with a metal mask for sample CA1and ultra-violet (UV) lithography photoresist as mask for CA2 and CA3. The exposed area is etched with argon plasma. The UV photoresist is cleaned with acetone and deionized water. All measurements were carried out at low temperatures down to 2K with a magnetic field up to 9T. Standard lock-in amplifiers (Stanford Research 830 and 850) were used to acquire data with the electric current of 1μA. The parameters of all samples are listed in **Table I**.

**Acknowledgements**

We gratefully acknowledge the financial support of the National Key Projects for Basic Research of China (Grant Nos: 2013CB922103, 2011CB922103), the National Natural Science Foundation of China (Grant Nos: 91421109, 11574133, 61176088, and 11274003), the PAPD project, the Natural Science Foundation of Jiangsu Province (Grant BK20130054), and the Fundamental Research Funds for the Central Universities. We would also like to acknowledge the helpful assistance of the Nanofabrication and Characterization Center at the Physics College of Nanjing University.

**Author contributions**

B. Z., F. S. and F.X. conceived the work and wrote the paper. B. Z., F.X. and P. C. prepared the samples. B. Z., S. Z., H. P. and F. S. performed the experiments. F.X., B. W., F. S. and G. W. had valuable discussions and edited the manuscript. All authors commented on the manuscript.

**Additional information**

**Competing financial interests:** The authors declare no competing financial interests.

**Tables**

Table I. Measured and calculated $Cd_3As_2$ sample parameters. The device length is the distance between the two measure voltage probes in a four-probe configuration. The values of $R_{xx}, R_H, n_e, \mu, \ell_e$ are for 2K. The longitudinal resistance $R_{xx}$, Hall coefficient $R_H$ obtained from fitting the Hall resistance with linear curve. The carrier density $n_e = 1/eR_H$, the mobility $\mu$ obtained from $\sigma = 1/\rho = (L/W)1/R_{xx} = n_e e \mu$, the mean-free path $\ell_e = \hbar\mu\sqrt{2\pi n_e}/e$.

| Device | $L$(μm) | $W$(μm) | $R_{xx}(\Omega)$ | $R_H(\Omega/T)$ | $n_e(10^{12}cm^{-2})$ | $\mu$ ($cm^2/Vs$) | $\ell_e$(nm) |
|---|---|---|---|---|---|---|---|
| CA1 | 30 | 80 | 416.8 | 290.36 | 2.15 | 2612.4 | 64 |
| CA2 | 25 | 25 | 4039.8 | 991.42 | 0.63 | 2454.1 | 32 |
| CA3 | 4 | 4 | 1520.2 | 377.1 | 1.66 | 2480.5 | 53 |

**Figures**

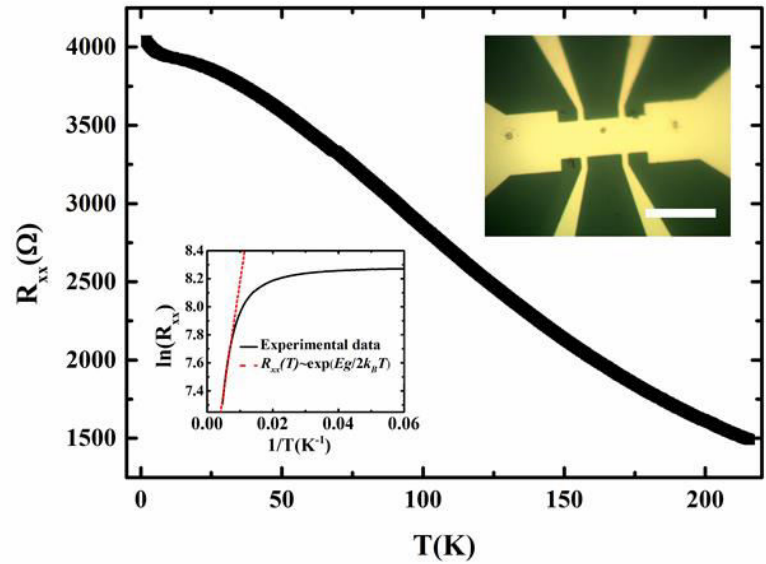

**Figure 1 |** Temperature dependence of the resistance of the $Cd_3As_2$ film device CA2. The right inset shows its optical image with the scale bar of 50 µm. The left inset shows the Arrhenius fitting of $R_{xx}(T) \sim exp(E_g/2k_BT)$ with the result of a 21.9meV band gap.

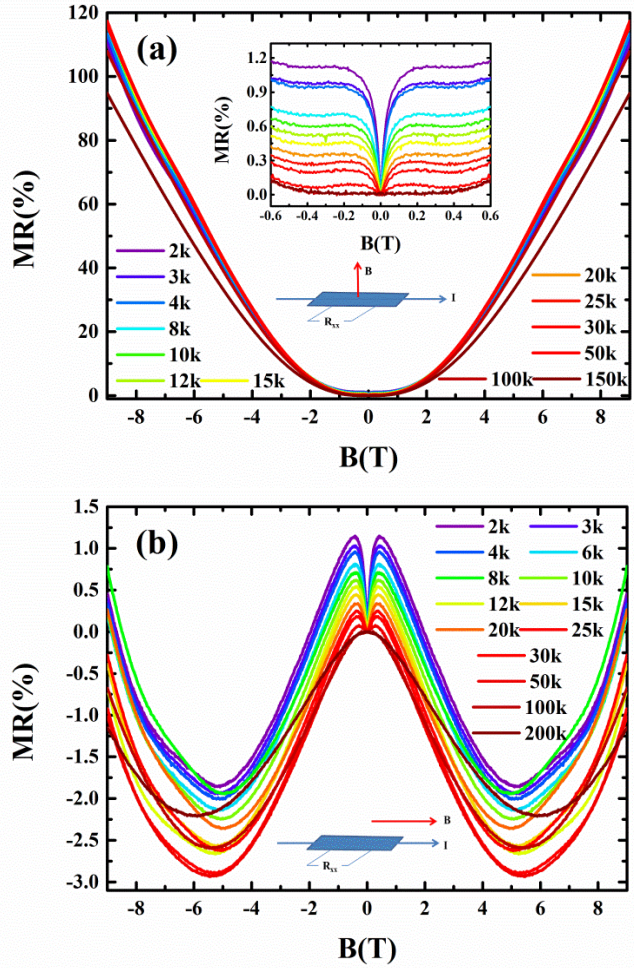

**Figure 2 |** The MR with the applied magnetic field $B \perp E$ or $B \mathbin{/\!/} E$. Schematic diagram of electrical transport measurements are shown in insert respectively. **(a)** The magnetoresistance vs magnetic field in $B \perp E$ at various temperatures from 2K to 150K. The insert show the sharp cusp at low magnetic field. **(b)** The magnetoresistance with $B \mathbin{/\!/} E$ at various temperatures from 2K to 200K. The WAL effect is also clear around zero fields. With the magnetic field increasing, the MR decreasing shows a negative MR phenomenon with a weak temperature dependent below 150K. However, it increase with increasing magnetic field when B>5T.

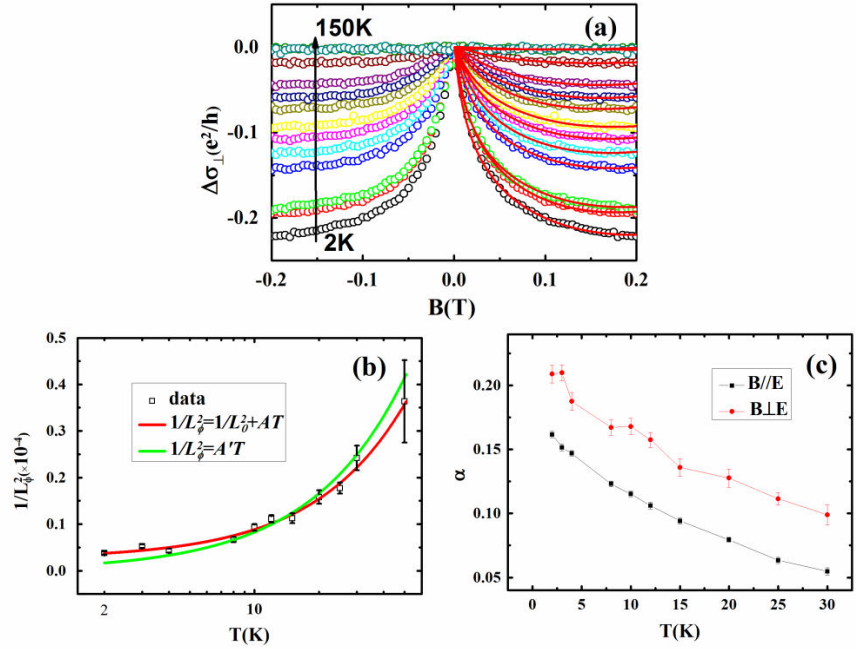

**Figure 3 |** Weak antilocalization effect in applied magnetic field B ⊥ E. **(a)** The magnetoconductivity of sample CA1 (hole circle) vs magnetic field in B⊥E at temperature from 2K to 150K along with fitting (red solid curves) to the Equation (1). **(b)** The dephasing length vs temperature. **(c)** Temperature dependent parameter α with B⊥E or B∥E.

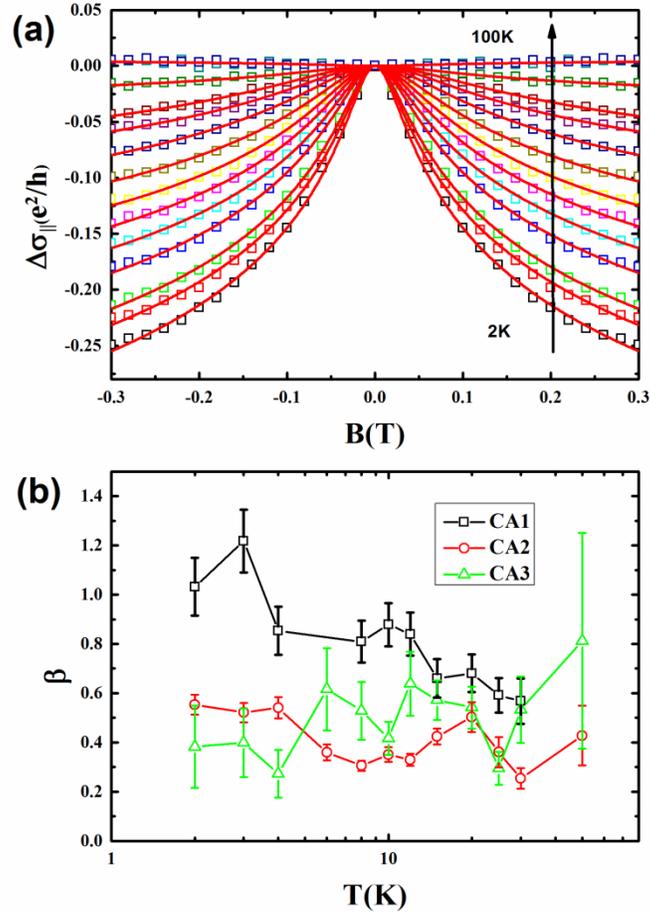

**Figure 4 |** Weak antilocalization effect in applied magnetic field B ∥ E. **(a)** The magnetoconductivity of sample CA1 (hole square) vs magnetic field in B ∥ E at temperature from 2K to 100K along with fitting (red solid curves) to the Equation (2). **(b)** Temperature dependence of β obtained from equation (2). For sample CA1, the mean free path 64nm is larger than the film thickness 50nm.